\documentclass[11pt]{article}
\pdfoutput=1
\usepackage{tikz}
\usetikzlibrary{decorations.pathmorphing}
\usepackage{putex}
\usepackage{feyn}
\usepackage[vcentermath]{youngtab}
\usepackage{subfig,braket}
\usepackage{lscape}
\usepackage{indentfirst}
\usepackage{graphicx}
\usepackage{epstopdf}
\usepackage{enumerate}
\usepackage{cite}
\usepackage{tensor}
\usepackage{slashed}
\usepackage{amsmath}
\usepackage{amssymb}
\usepackage{mathrsfs}
\usepackage{lgrind}
\newcommand{\be}{\begin{equation}}
\newcommand{\bea}{\begin{eqnarray}}
\usepackage{authblk}
\usepackage[font=footnotesize,labelfont=bf]{caption}

\usepackage{bbm}

\usepackage{hyperref}

\numberwithin{equation}{section}

\newcommand {\bes} {\begin {equation*}}
\newcommand {\ees} {\end {equation*}}

\def\CH{{\cal H}}

\def\be{ \begin{equation} }
\def\ee{ \end{equation} }

\def \beta {\beta}

\def \beq { \begin{equation}}
\def \eeq {\end{equation}}

\usepackage[normalem]{ulem}

\begin{document}

\author[1]{Stefano Negro}
\author[1]{Fedor K. Popov}
\author[2]{Jacob Sonnenschein}

\affil[1]{CCPP, Department of Physics, NYU, New York, NY,  10003, USA}
\affil[2]{The Raymond and Beverly Sackler School of Physics and Astronomy, \\
Tel Aviv University, Ramat Aviv 69978, Tel Aviv, Israel}
\renewcommand{\Affilfont}{\small\it}
\title{Deterministic Chaos vs Integrable Models}
\date{\vspace{-5ex}}
\maketitle

\abstract{
In this work we present analytical and numerical evidences that classical integrable models possessing infinitely many degrees of freedom unexpectedly exhibit some features that are typical of chaotic systems. By studying how the conserved charges change under a small deformation of the initial conditions, we conclude that the inverse scattering map is responsible for the presence of these features, in spite of the system being integrable. We investigate this phenomenon in the explicit examples of the KdV equation and the sine-Gordon model and further provide general arguments supporting this statement.}
\section{Introduction}
Our understanding of physics consistently advanced thanks to the study of exactly solvable models -- from the Kepler problem \cite{babelon2003introduction}, to the statistical mechanics of lattice systems \cite{baxter2016exactly}, to the latest advances in interacting quantum field theories (QFTs). A powerful illustration of this approach is given by integrable QFTs (IQFTs) in two spacetime dimensions; see, e.g., refs.  \cite{Dorey:1996gd, Bombardelli:2016rwb} for reviews. These systems possess infinitely many independent, mutually commuting conserved quantities \cite{Bazhanov:1994ft,Bazhanov:1996aq,Bazhanov:1996dr,Negro:2016yuu}. Their existence constrains the dynamics to the point of allowing the efficient computation of a wealth of observables -- something very remarkable for an interacting QFT. Another interesting  development of modern physics has been  the understanding of the phenomena of chaos and thermalization \cite{maldacena2016remarks,maldacena2016bound,gross2021chaotic,deutsch1991quantum,Motamarri:2021zwf}. These last two terms are usually assumed to be interconnected and the corresponding concepts are often conflated, implying that thermalization necessarily entails chaos and chaos inevitably leads to the thermalization. As a consequence, it is commonly believed that integrable models are incompatible with chaos, since the large amount of conservation laws prevents the system from thermalizing. In this letter, we will present evidence of what appears to be chaotic behavior in integrable models. 

Integrable systems come in various forms, the simplest of these being the exactly solvable models of classical mechanics, whose phase space is finite-dimensional \cite{babelon2003introduction}. Due to this last property one can  prove that these systems are indeed incompatible with the concept of chaos. However this  proof hinges on the fact that the dimension of the phase space is finite, and therefore cannot be applied to integrable theories on a general ground. An intriguing question is then whether such a statement applies to integrable systems with an infinite-dimensional phase space. In this note we will present evidence suggesting that the answer to this question might not be a straightforward yes, as one might naively expect. Before diving right in, we wish to clarify the notions of \emph{integrability} and \emph{chaos} we are going to employ, in order for the significance of our findings and statements to be clear. We will use the following definitions
\begin{enumerate}
    \item {\it Integrability}. We refer to a system as \emph{integrable} if there exists a method that allows one to solve the theory -- \emph{i.e.} find all physical observables -- with a finite number of quadradures \cite{babelon2003introduction} or algebraic manipulations. 
    \item {\it Chaos}. We will say that a system is \emph{chaotic} -- equivalently, that it exhibit \emph{deterministic chaos} -- if, for any initial state, one can find a small deformation that drives a system away under time evolution at least in a weak sense, namely that the deviation grows and is unbounded \cite{guckenheimer2013nonlinear,PlatoStanford}.We can state this in the following way
    \begin{gather}
        \forall C,\delta>0 \quad \forall x(0) \in \mathbb{M} \quad \exists \left|y(0)-x(0)\right|<\delta  \exists t>0: \left|y(t)-x(t)\right| > C\,,
    \end{gather}
    where $\mathbb{M}$ is ourconfigurationspace and we assume that $x(t)$ is the trajectory of our system as a function of time $t$.
\end{enumerate}
Let us elaborate on these definitions. The above notion of chaos implies that if we know the initial state of our system only with some finite precision, then the precision with which we can predict the state of the system will deteriorate as we let the latter evolve in time. Stated differently, any error in the determination of initial state will become arbitrarily large after the system has evolved for a sufficient amount of time. This feature is known as the \emph{butterfly effect} or \emph{weak butterfly effect}, when we have just a growth instead of exponential one.

For what concerns integrability, its definition is a notoriously slippery one \cite{zakharov1991integrability, hitchin2013integrable}, which is often used interchangeably with the concept of \emph{exact solvability}, though the two are not necessarily synonymous \cite{Torrielli:2016ufi}. We will not address this matter, but will content ourselves with the above working definition. This is purposefully broad to include the exactly solvable models, the standard Liouville integrable systems and integrable Partial Differential Equations (PDEs) that will be considered in the following. An example of a theory where the concepts of integrability and chaos appear to coexist is the free quantum field theory in Rindler space. This theory is clearly integrable while still exhibiting some thermal properties and a non-vanishing Lyapunov exponent \cite{zheng2003observer}. We thus have at least one case in which integrability and chaos do not necessarily exclude each other.
One last definition that we will need in the following is that of unstable map which, as we will remark momentarily, is deeply related with the concept of chaos as defined above.
\begin{itemize}
    \item[3] \emph{Unstable maps}. We will say that a map $f$ between two metric spaces $(d_1,M_1)$ and $(d_2,M_2)$ is \emph{unstable} if there exist nearby points in $M_1$ that can be sent to distant points in $M_2$. More rigorously $$\forall \epsilon>0\;,\; M>0\;,\quad \exists x,y \in M_1 \;:\; d_1(x,y) <\epsilon\;,\;  d_2\Big(f(x),f(y)\Big)>M\;.$$
\end{itemize}
Clearly, here there is a strong dependence on the definition of \emph{distance} employed. For example, a map inducing a metric on the target space will not exhibit the butterfly effect. However, the vast majority of physical problems comes equipped with a natural measure, making the concept of unstable map presented above a relevant one. Note that for smooth, finite dimensional spaces $M_1,M_2$, a smooth map $f$ cannot be unstable. For infinite-dimensional spaces, on the other hand, this is what seems to happen in general.

Before proceeding with our analysis, we wish to briefly explain in simple terms why the concept of unstable map is important. The usual procedure to solve the dynamics of a classical integrable theory is to exploit a map from the original set of coordinates to a new one in which the time evolution of the system is very simple -- oftentimes linear. For example, in the case Liouville integrable systems, one reduces a complex dynamics to a set of independent harmonic oscillators by employing the so-called \emph{action-angle variables} \cite{Torrielli:2016ufi}. One then evolves the system in the latter coordinate system and, finally uses the same map in reverse to return to the starting, physical coordinate description. Were such a map between coordinate systems to be unstable in the sense above, then any small error in the determination of initial conditions might grow arbitrarily large when employing the coordinate map and its inverse. What this implies is that if we start with nearby initial conditions, the time evolution, as determined by the procedure described here, might bring them arbitrarily far apart. This is precisely the definition of deterministic chaos we have given above. As we are going to see in this work, this situation arises when the phase space of the integrable system is infinite dimensional. Let us stress that we are not claiming that these integrable systems are inherently chaotic, merely that the standard methods for the solution of their dynamics may produce, at a computational level, behaviours typically associated with chaotic systems.

The paper is organized as follows. In the next Section \ref{sec:din_sys} we begin investigating how chaotic features can appear in integrable systems. We first discuss theories with finite dimensional phase spaces. We will see that the differentiability of the latter is a pivotal criterion for the absence of chaotic behavior. Next we consider two examples of exactly solvable discrete-time dynamical systems that exhibit deterministic chaos. Their simplicity will allow us to pinpoint the source of such chaotic behavior. We will see that this arises from a map between the position variables and a set of auxiliary ones in which the dynamics of the systems becomes, in a sense, linear. Section \ref{sec:IFT} is devoted to integrable systems with an infinite dimensional phase spaces: integrable field theories or integrable PDEs. We will consider two famous examples: the KdV equation and the sine-Gordon model. Here too we will see that there is chaotic behavior emerging from a map -- the \emph{inverse scattering transform}\cite{zakharov1980inverse, faddeev1987hamiltonian} -- between two sets of variables, in this case the field variables on one side and the conserved charges on the other. We relate the unstable nature of this map to the fact that the propagator of free particles in $1$ dimension is IR-divergent. This causes the inverse scattering map to be very sensitive to small perturbations. We conclude and present a number of open questions in Section \ref{sec:conc_out}. We tried to keep the content of this letter accessible to non-specialized readers and dedicated the appendices for the more technical aspects. In Appendix \ref{app:pert_scatt} we briefly present a general result concerning the instability of the inverse scattering transform under small variations of the initial conditions. Appendix \ref{app:mon_mat} is devoted to a lightning review of the procedure to compute the conserved charges in classical integrable field theories.
\section{Integrability and Chaos in Dynamical Systems}
\label{sec:din_sys}
\subsection{Finite Dimensional Integrable Systems}\label{subsec:FDIS}
We start from the case of finite dimensional integrable systems. Specifically, let us first consider models whose phase space is a differentiable manifold. Then we can conclude that their dynamics cannot be chaotic. Indeed, let us fix the dimension of the phase space to be $2n$. The system being integrable, it is possible to find at least $n$ algebraically independent conserved charges $I_j$ in involution: $\left\{I_i, I_j \right\} = 0$. Consequently, there exist a canonical map to action-angle variables in terms of which the dynamics takes place on an $n$-torus \cite{babelon2003introduction}. Since this map is finite-dimensional and, at generic points, smooth it cannot generate arbitrarily large deviations from small ones. Therefore the map is not unstable. We conclude that finite dimensional integrable systems whose phase space is a differentiable manifold cannot exhibit deterministic chaos.

\begin{figure}
    \centering
    \includegraphics[scale=.4]{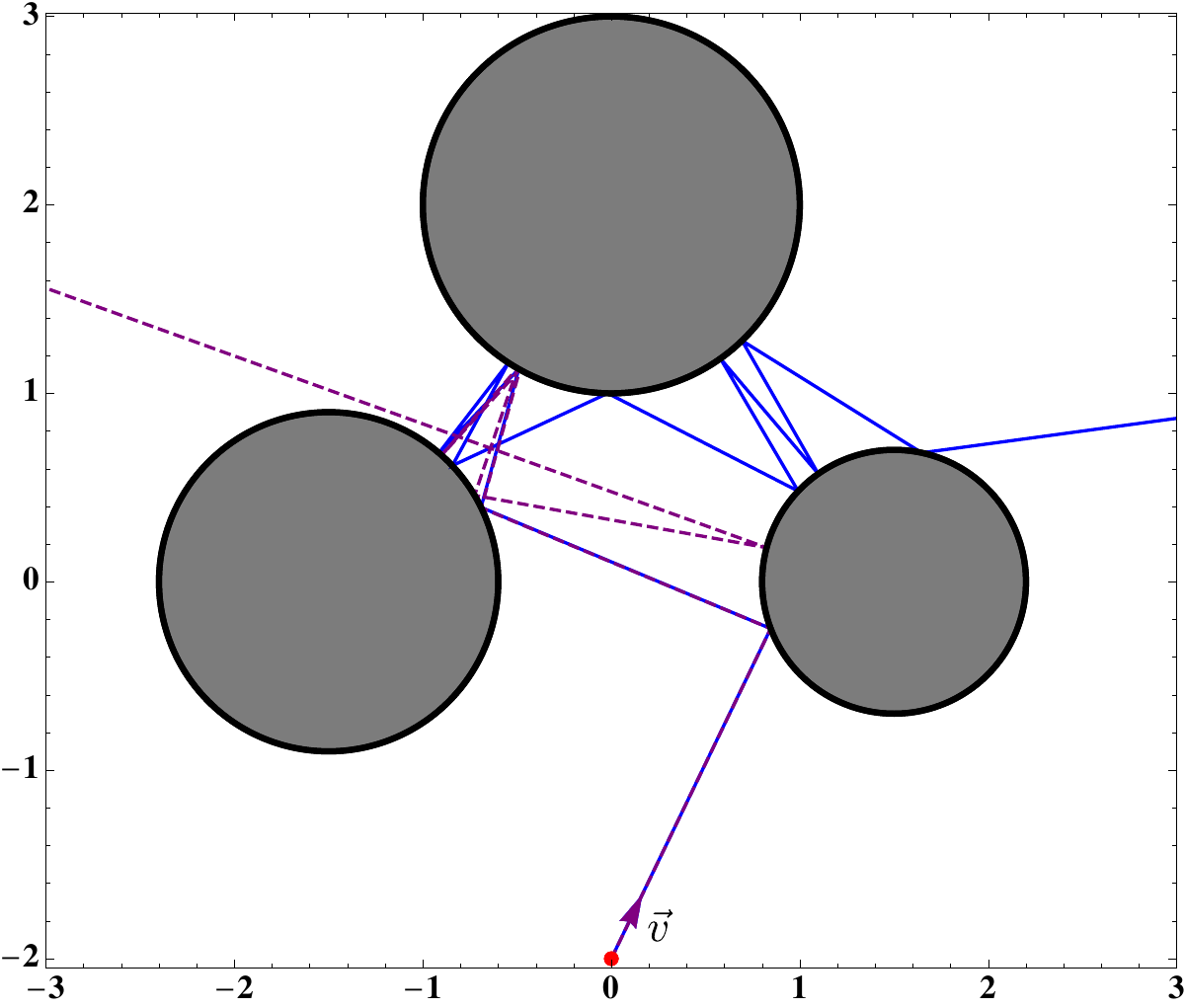}
    \caption{Two trajectories of a pin-ball scattering, issuing with the same velocity $\vec{v}$ from nearby initial positions $x=(0,-2)$ (the blue, full line) and $x'=(0,-2-5\cdot 10^{-4})$ (the purple, dashed line). The trajectories are indistinguishable at first, begin diverging after a few scatterings, and end up in completely different regions for late time.}
    \label{fig:pinball}
\end{figure}

This argument relies importantly on two points: the finite dimensionality and the differentiability of the phase space manifold. If we relax the latter, it becomes easy to find a counter-example to the incompatibility of chaos and integrability.
In fact, let us consider the \emph{pin-ball problem}: a two dimensional system of a point particle scattering rigidly against three disks (see fig. \ref{fig:pinball}).
This problem obeys our definition of integrability: from any initial position and velocity it is possible to determine the position at any later time through a finite number of algebraic manipulations. Indeed, in-between the collisions the particle moves freely and at the collision points we know exactly what happens: the velocity of the particle gets reflected with respect to the plane perpendicular to the circle at the collision point. However it is very well-known that this  system is chaotic \cite{smilansky1991classical}. A small change of initial position and velocity vectors can drastically change the trajectory at later times. This simple system has even been observed to exhibit some kind of fractal behavior \cite{eckhardt1987fractal}. The reason for the appearance of chaotic behavior, even though the system is finite-dimensional, is to be found in the non-differentiability of the phase space manifold.

\subsection{Deterministic Chaos in Discrete-Time Dynamical Systems}
The other way in which the argument we presented above can fail is if the phase space of the system is infinite dimensional. In this case a smooth map can indeed send small variations to large ones. Let us then investigate more in detail infinite dimensional systems. As a warm up, we are going to first consider some toy examples so that we can appreciate in a simple setting how chaos manifests itself. The first example is the Bernoulli dynamical system \cite{guckenheimer2013nonlinear}, defined as follows
\begin{equation}
    R\;:\;\begin{array}{l c l}
        \left[0,1\right) & \longrightarrow & \left[0,1\right) \\
        x & \longmapsto & R(x) = (2x) \mod 1
    \end{array}\;.
\end{equation}
This system is integrable. Indeed any real number $x\in \left[0,1\right)$ can be mapped into an infinite binary sequence $\lbrace\sigma_i\rbrace_{i=1}^{\infty}$:
\begin{equation}
    x = \sum^\infty_{i=1} \frac{\sigma_i}{2^i}\;.
\end{equation}
The Bernoulli map $R(x)$ acts on the binary sequence by shifting it to the left $\sigma_i \to \sigma_{i+1}$ and removing $\sigma_0$ from the set. So from a given initial condition $x_0$ we can determine the future state $R^n(x)$ for any finite $n$ performing a finite number of manipulations on the sequence $\lbrace\sigma_i\rbrace_{i=1}^{\infty}$. Nonetheless, the system clearly presents a chaotic behavior. Indeed, a rational initial condition $x_0$ yields a periodic trajectory\footnote{If $x_0$ is an inverse power of $2$, the trajectory actually collapses to the fixed point $x=0$ after a finite amount of iterations.}, i.e. $x_0$ is the fixed point of the $n$-th iterate of $R$ for some positive integer $n$: $R^{n}(x_0)=x_0$. However an irrational $x_0$ generates an aperiodic trajectory. Now, since the sets of rational and irrational numbers are dense we conclude that a slight change of the initial conditions can drastically change the behavior of the system. Notice that the binary representations of rational and irrational numbers differ vastly: rational numbers possess a finite or recurring infinite binary representation, while for irrational numbers the binary representation is aperiodic. In other words, altering slightly a number $x$ greatly affects its binary representation and we can say that the map $x\to \left\lbrace\sigma_i\right\rbrace$ is unstable. This is the basic reason underlying the chaotic properties of the Bernoulli map and we will see that the same principle is at play in general integrable systems.

In order to get closer to the systems that possess an infinite number of conserved charges, we can consider the Baker's map, defined in the following way
\begin{equation}
    B\;:\;\begin{array}{l c l}
        \left[0,1\right)^2 & \longrightarrow & \left[0,1\right)^2 \\
        \left( x, y \right) & \longmapsto & B(x,y) = \left\lbrace \begin{array}{l r} \left(2x, \frac{y}{2}\right)\;, & x \in \left[0,\frac{1}{2}\right) \\ \left(2-2x, 1-\frac{y}{2}\right) \;, & x \in \left[\frac{1}{2},1\right) \end{array} \right.
    \end{array}\;.
\end{equation}
This map is integrable as well: as before we can map the position variables $(x,y)$ to a binary sequence, this time doubly-infinite
 \begin{equation}
    I\;:\;\begin{array}{l c l}
        \left[0,1\right)^{2} & \longrightarrow & \left\lbrace0,1\right\rbrace^{\mathbb Z} \\
        (x,y) & \longmapsto & \left\lbrace\sigma_i\right\rbrace_{i=-\infty}^{\infty}\;.
    \end{array}
 \end{equation}
 where
\begin{equation}
    x = \sum_{i=0}^{\infty} \frac{\sigma_{-i}}{2^{i+1}}\;,\qquad y = \sum_{i=0}^{\infty}\frac{\sigma_{i+1}}{2^{i+1}}\;.
\end{equation}
Then the composition of this map with the Baker's one simply acts as a left shift of the binary sequence
\begin{equation}
    I\circ B\circ I^{-1}\;:\;\left\lbrace\sigma_i\right\rbrace_{i=-\infty}^{\infty}\;\longmapsto\;\left\lbrace\tilde{\sigma}_i = \sigma_{i+1}\right\rbrace_{i=-\infty}^{\infty}\;.
\end{equation}
In this system, any totally symmetric function of the $\sigma_i$ is a conserved quantity, e.g.
\begin{equation}
    \mathfrak{q}_{2n+1} = \lim_{N\rightarrow\infty}\frac{1}{2N+1} \sum_{j=-N}^{N} \prod_{k=-n}^n s_{j+k}\;,\qquad s_j = \sigma_j - \frac{1}{2}
\end{equation}
However, just as for the Bernoulli map, the behavior of the trajectories is chaotic. We see that the cause is the same as for the Bernoulli map. When described in the binary sequence ``system of coordinates'', the Baker's map is not chaotic: any small change in the initial binary sequence will remain small in the subsequent dynamics. The chaotic behavior appears in the translation from the position variables $(x,y)$ to the binary sequence $\left\{\sigma_i\right\}^\infty_{n=-\infty}$. A small deviation in $x$ can cause an arbitrary large change in the binary sequence which, in turn, produces a large difference in the dynamics. One might argue that it is entirely possible to perform all the necessary calculations to determine the trajectory at any point in the future using only the binary digits ``coordinates'' and, consequently, the system is inherently non-chaotic. However, let us suppose, for the sake of the argument, that we are using the Baker's map to model a real-world physical system. In this case we might want to make predictions on the position of the system as a function of time and of its initial position. Then we cannot ignore the fact that trajectories issuing from nearby initial conditions can diverge after a finite amount of time. In this perspective, the position variables $(x,y)$ have a clear physical interpretation, while the binary digits are a mathematical construct, an helpful tool that solves the specific model we are dealing with. Mapping from the latter to the former can produce chaotic behavior in the system.

As we will see later, this feature is also present in integrable PDEs, such as the KdV equation and the sine-Gordon model that we will consider below. Before moving to these cases, we wish to consider another example where integrability and chaos seem to coexist: the problem of reconstructing an analytic function from its values on a compact set. Take some function $\alpha(t)$ defined on the unit interval $t\in \left[0,1\right]$. We wish to find a function $f(z)$, analytic in the whole complex plane $z\in\mathbb{C}$ and such that
\begin{equation}
    f(t) = \alpha(t), \qquad t\in \left[0,1\right]\;.
\end{equation}
This problem is integrable: a solution exists, is unique and can be constructed by a finite number of integrations. It is also easy to see that it is chaotic. Indeed, let us introduce a small change in the ``initial conditions'' $\alpha(t)$ as follows
\begin{equation}
    \alpha(t)\;\longrightarrow\;\alpha'(t) = \alpha(t) + \frac{C}{t^2+1}, \quad C \ll 1\;.
\end{equation}
The deformation $\delta\alpha = \alpha' - \alpha$ is smaller or equal to $C$ everywhere in the interval $t\in[0,1]$ where the initial conditions are defined. However the analytic function $f(z)$ receives arbitrarily large corrections in the vicinity of $t= \pm i$. In other words, this problem is unstable or chaotic for small deformations.
\section{Chaotic Behavior in Integrable Field Theories}
\label{sec:IFT}
\subsection{The KdV Equation}
Now let us move to classical integrable systems with infinite dimensional phase space, that is, integrable field theories or integrable PDEs \cite{zakharov1980inverse}. We will start with one of the most renowned examples: the Korteweg–de Vries (KdV) equation
\begin{equation}
    \partial_\tau u(x,\tau) + 6 u(x,\tau) \partial_x u(x,\tau) + \partial_x^3 u(x,\tau) = 0\;.
\label{eq:KdVeq}
\end{equation}
This system is famously integrable and can be solved by means of the \emph{inverse scattering transform}\cite{zakharov1980inverse, faddeev1987hamiltonian}. Namely, let us consider the following scattering problem
\begin{align}
    &\Big(-\partial_x^2 + u(x)\Big) \psi_k(x) = k^2 \psi_k(x), \notag\\
    &\psi_k(x) \sim \left\{
    \begin{array}{l l}
    e^{-ik x}  & x \to -\infty \\
   \frac{1}{t_k} e^{-i k x} + \frac{r_k}{t_k} e^{i k x} & x \to +\infty
    \end{array}
    \right.\;,
\label{eq:KdVscattering}
\end{align}
where we take the potential $u(x)$ to correspond to the initial condition for the KdV equation: $u(x) = u(x,\tau = 0)$. Then, if we demand that $u(x,\tau)$ evolves in time according to the KdV equation, the scattering data -- that is the transmission and reflection coefficients $t_k$ and $r_k$ --  will have the following simple time dependence
\begin{equation}
    r_k(\tau) = r_k(0) e^{8 i k^3 \tau}\;,\quad t_k(\tau) = t_k(0)\;.
\end{equation}
For the case of a bound state, the scattering problem reads
\begin{align}
    &\Big(-\partial_x^2 + u(x)\Big) \psi_k(x) = -\kappa_n^2 \psi_k(x), \notag\\
    &\psi_n(x) \sim \left\{
    \begin{array}{l l}
    e^{\kappa_n x} & x \to -\infty, \\
    b_n e^{- \kappa_n x} & x \to +\infty
    \end{array}
    \right. \;,
\label{eq:KdVscatteringbound}
\end{align}
and the coefficients $b_n$ will also have a simple time dependence:
\begin{equation}
    b_n(t) = b_n(0) e^{8 \kappa_n^3 t}\;.
\end{equation}
Now, the task of the inverse scattering method is to reconstruct the potential $u(x,\tau)$ from the scattering data $t_k(\tau)$, $r_k(\tau)$ and $b_n(\tau)$. This can be achieved thanks to the \emph{Gel'fand-Levitan-Marchenko} integral equation \cite{gel1951determination, marchenko2011sturm}. So the inverse scattering method acts as a ``non-linear Fourier transform'' of sorts, mapping the coordinates $u(x,\tau)$ to the scattering data $t_k(\tau)$, $r_k(\tau)$, $\kappa_n$ and $b_n(\tau)$ in terms of which the time evolution is almost trivial. However this map is unstable, meaning that a small perturbation in the scattering data can generate arbitrarily large deviations in the reconstructed potential \cite{dorren1994stability,carrion1986stability,feinberg2004response}. This is a very well-known fact in geophysics where the inverse scattering method is used to reconstruct the internal structure of a medium from the back-scattered signal  \cite{dorren1994stability,berryman1980discrete}.

The direct scattering problem -- that is the process of passing from the non-linear equation \eqref{eq:KdVeq} to the scattering representation (\ref{eq:KdVscattering}, \ref{eq:KdVscatteringbound}) -- also turns out to be unstable. Indeed, it is a well-known phenomena that in $1$ and $2$ space dimension any small dip or well in $u(x)$ must correspond to a bound state \cite{landau2013quantum}. Therefore a small change in $u(x)$ or in $u'(x)$ might create bound states and drastically alters the scattering data $t_k$, $r_k$, $\kappa_n$ and $b_n$. Let us stress the fact that not all small variations of the potential $u(x)$ are bound to cause drastic changes in the scattering data. For instance, if we change $u(x)$ by letting it evolve with respect to the KdV equation, the scattering data will only changes slightly. The important point is that the overwhelming majority of changes in the initial conditions modifies considerably the future behavior of the system.

The reason for the presence of such a behavior typical of chaotic systems, from the perspective of the scattering problem, can be traced back to the fact that in $1$ and $2$ dimensions the latter is ``strongly-coupled''. Thus perturbation theory cannot be employed. Indeed, let us consider again the scattering problem \eqref{eq:KdVscattering} and take a very small potential $u(x)$ that we can consider as a small deformation of zero. At first order in perturbation theory we can write 
\begin{equation}
    \psi_k(x) = e^{-i k x} + \delta \psi(x)\;, \quad \left(\partial_x^2 + k^2\right) \delta \psi(x) = u(x) e^{-i k x}\;.
\end{equation}
The solution of this equation is 
 \begin{equation}
     \delta \psi(x) = \int \frac{dy}{4\pi i} \frac{e^{ik|x-y|}}{k} u(y) e^{-ik y},
 \end{equation}
and, as one immediately sees, in the vicinity of $k=0$ this correction can be become arbitrarily large, meaning perturbation theory breaks down. While the reconstruction of the potential is not sensitive to changes in the UV range of the scattering data, the IR one can drastically modify the resulting potential. We show this fact more in general in the Appendix \ref{app:pert_scatt}. Let us stress that this instability occurs only in $1$ and $2$ dimensions, where the propagator of a free particle diverges at large distances.

Notice how the situation described above has some resemblance to the one of the Bernoulli and Baker's maps mentioned in the previous section. Here as well we have a map from the field variables $u(x)$ and $\partial_x u(x)$ -- the ones in which the initial conditions of the system are given -- to a set of auxiliary variables in which the dynamics is easily solvable -- the scattering data. As it was the case in the previous examples, this map turns out to be unstable. We can gain a more concrete feel of this instability, by computing numerically the conserved charges of the KdV equation for two very similar initial profiles. An expression of the conserved charges can be obtained using the following recursive relation \cite{zakharov1980inverse} (see also Appendix \ref{app:mon_mat}):
\begin{equation}
    Q_n = \intop_{-\infty}^{\infty} w_{n-1}(x) dx\;,\qquad \left\lbrace \begin{array}{l}
        w_0(x) = u(x) \\
        w_1(x) = \partial_x u(x) \\
        w_n(x) = \partial_x w_{n-1}(x) + \sum\limits^{n-2}_{k=0} w_k(x) w_{n-2-k}(x)
    \end{array}\right.\;.
\end{equation}
It is possible to check that $Q_{2n} = 0$ -- since $w_{2n-1}$ are total derivatives. We numerically evaluated the first few low-lying charges for the trivial solution $u(x) = 0$ and for a small perturbation in the form of a Gaussian distribution
\begin{equation}
    \tilde{u}(x) = 0.1 \exp\left(-x^2\right)    
\end{equation}
The charges for the trivial solutions are obviously all vanishing
\begin{equation}
    Q_{2n-1} = 0\;,\quad \forall n\geq1\;,
\end{equation}
while the numerical values of the charges for $\tilde{u}(x)$ are
\begin{align}
    & \tilde{Q}_1 = 1.772\cdot 10^{-1}\;,\quad \tilde{Q}_3 = 1.253\cdot 10^{-2}\;,\quad \tilde{Q}_5 = -1.049\cdot 10^{-2}\;,\quad \tilde{Q}_7 = 3.122\cdot 10^{-2}\;, \notag \\
    & \tilde{Q}_9 = -1.527\cdot 10^{-1}\;, \quad \tilde{Q}_{11} = 1.041 \;,\quad \tilde{Q}_{13}= -9.066 \;, \quad \tilde{Q}_{15} = 9.582\cdot 10^1 \;, \\
    & \tilde{Q}_{17} = -1.186 \cdot 10^{3}\;,\quad \tilde{Q}_{19} = 1.6804 \cdot 10^{4}\;,\quad \tilde{Q}_{21} = 2.670\cdot 10^{5} \;,\quad \ldots \;. \notag
\end{align}
While the magnitude of first $5$ charges remains of the order of or smaller than $10^{-1}$, we see that as we increase the index, the charges grow larger and larger. This suggests that the map from initial profile $u(x)$ to the data of conserved charges is indeed unstable: a perturbation by a small Gaussian distribution leads to a drastic change of the charges.
\subsection{The sine-Gordon Model}
\label{sec:sG_model}
The previous considerations also apply to other integrable PDEs. Here we consider another famous example: the sine-Gordon model. This model can also be solved with the inverse scattering method. Therefore, due to the instability of the inverse scattering problem, we should observe similar properties as the one witnessed above. In particular we expect an arbitrary magnification of the errors in the map from the initial configuration to the space of conserved charges. While the phenomenon studied here pertains purely to the classical theory, we allow ourselves to make some speculation on its quantum version. The quantum sine-Gordon model has a factorized S-matrix describing the scattering between solitons. While we are not claiming that this $S$-matrix is chaotic, we propose that the map from a given initial quantum state to the basis of states with defined soliton number is indeed unstable. In other words, we expect that a slight deformation of the initial state might lead to the appearance of an arbitrarily large number of additional solitons. These will participate in the scattering process, and thus drastically change its outcome.

Let us play the same game we did for the KdV equation and evaluate numerically the first few conserved charges for slightly different initial conditions. We will consider the sine-Gordon equation in the ``light-cone'' form
\begin{equation}
    \partial_+ \partial_- \phi(x_+,x_-) = - \sin \phi(x_+,x_-)\;,
\label{eq:sG_equation}
\end{equation}
and interpret $x_+$ as the time and $x_-$ as the spatial coordinates. The conserved charges for this equation can be computed following a recursive procedure \cite{candu2013introduction} similar to the one we presented above for the KdV equation (see also Appendix \ref{app:mon_mat})
\begin{equation}
    Q_n = \intop_{-\infty}^{\infty} p_n(x) dx\;,\qquad \left\lbrace \begin{array}{l}
        p_0(x) = \frac{i}{2}\partial_-\phi(0,x) \\
        p_1(x) = p^2_0(x) - \partial_- p_0(x) \\
        p_n(x) = -\partial_- p_{n-1}(x) - \sum\limits ^{n-2}_{k=1} p_k(x) p_{n-1-k}(x)
    \end{array}\right.\;.
\end{equation}
Here too it is possible to check that $Q_{2n} = 0$. First we are going to compare the charges associated to the stationary soliton
\begin{equation}
    \phi(0,x_-) = 4 \arctan e^{x_-}\;,
\end{equation}
and to the following, slightly altered profile
\begin{equation}
    \tilde{\phi}(x_-) = \pi\left(\tanh\frac{2x_-}{\pi} + 1\right)\;.
\end{equation}
\begin{figure}
    \centering
    \includegraphics{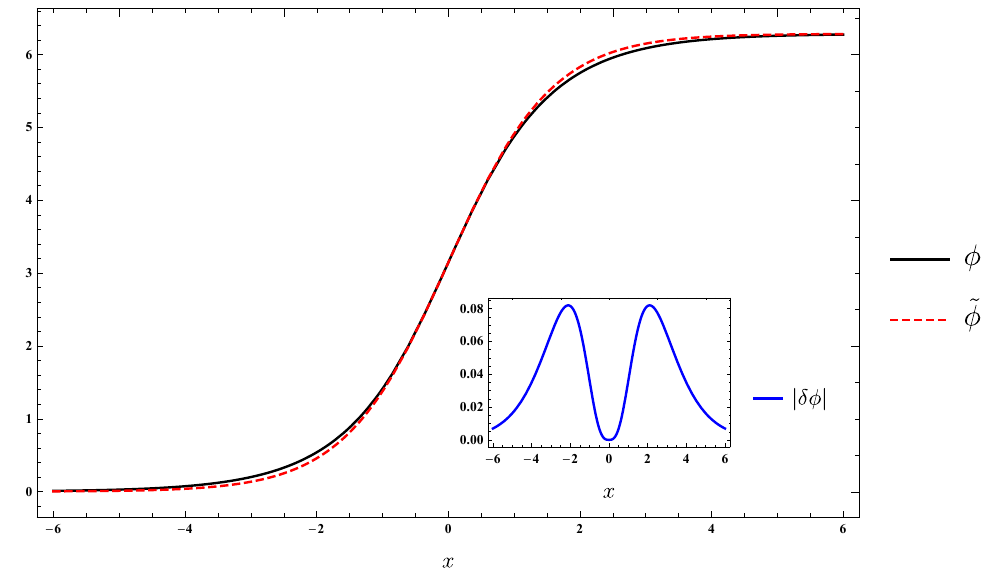}
    \caption{A plot of the two profiles $\phi(x) = 4\arctan e^{x}$ and $\tilde{\phi}(x) = \pi\left(\tanh\frac{2x_-}{\pi} + 1\right)$. The inset plot displays the absolute difference $\vert\delta\phi(x)\vert = \vert\tilde{\phi}(x) - \phi(x)\vert$.}
    \label{fig:sG_solitonVSalmost_solition}
\end{figure}
As one can see from fig. \ref{fig:sG_solitonVSalmost_solition}, these two profiles are quite similar for any real value of $x_-$. The charges for the soliton can actually be computed analytically to be inversely proportional to their index
\begin{equation}
    Q_{2n-1} = -\frac{2}{2n-1}\;.
\end{equation}
On the other hand, for the profile $\tilde{\phi}$ we resort to a numerical evaluation that yields
\begin{align}
    & \tilde{Q}_1 = -2.094\;,\quad \tilde{Q}_3 = -7.571\cdot 10^{-1}\;,\quad \tilde{Q}_5 = -5.205\cdot 10^{-1}\;,\quad \tilde{Q}_7 = -2.896\cdot 10^{-1}\;, \notag \\
    & \tilde{Q}_9 = -1.042\;, \quad \tilde{Q}_{11} = 6.168 \;,\quad \tilde{Q}_{13}= -7.461\cdot 10^{1} \;, \quad \tilde{Q}_{15} = 1.048\cdot 10^{3} \;, \\
    & \tilde{Q}_{17} = -1.775 \cdot 10^4 \;,\quad \tilde{Q}_{19} = 3.539 \cdot 10^5 \;,\quad \tilde{Q}_{21} = -8.182 \cdot 10^6 \;,\quad \ldots \;. \notag
\end{align}
The other pair of profiles we wish to compare are the trivial solution $\phi(x_-) = 0$ and the linear superposition of two solitons, separated by a small distance $2x_0$
\begin{equation}
    \tilde{\phi}(x_-) = 4 \arctan e^{x_- - x_0} - 4 \arctan e^{x_- + x_0}\;,\quad x_0\ll 1\;.
\end{equation}
\begin{figure}
    \centering
    \includegraphics{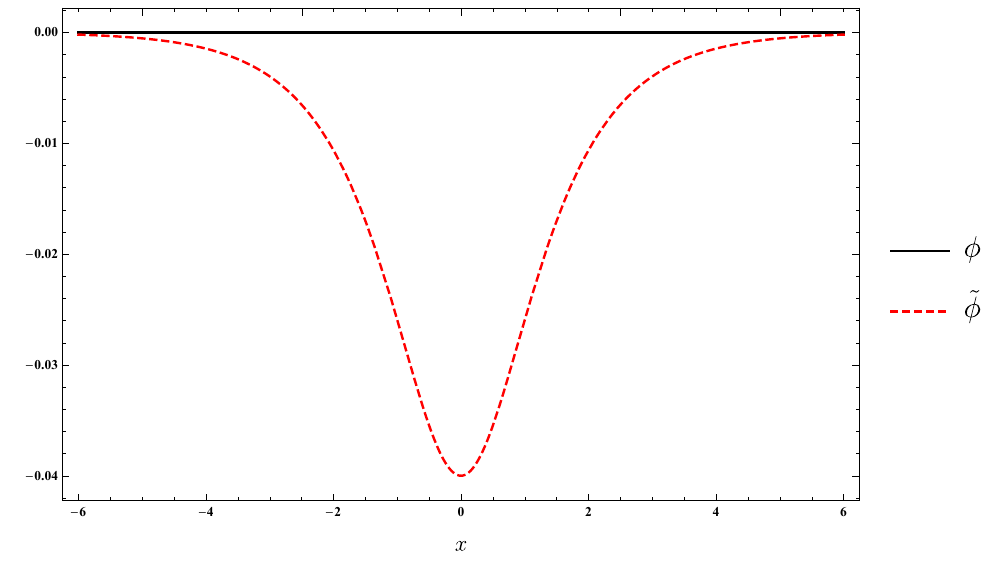}
    \caption{A plot of the profiles $\tilde{\phi}(x) = 4 \left(\arctan e^{x_- - 10^{-2}} - \arctan e^{x_- + 10^{-2}}\right)$ compared to the trivial function $\phi(x) = 0$.}
    \label{fig:sG_zeroVSalmost_zero}
\end{figure}
Figure \ref{fig:sG_zeroVSalmost_zero} shows that this profile differs from zero by the order of magnitude of $x_0$ and can thus be made arbitrarily small. The charges for $\phi(x_-) = 0$ are clearly all vanishing, while those for $\tilde{\phi}(x_-)$ are computed numerically for $x_0 = 10^{-2}$ to be
\begin{align}
    & \tilde{Q}_1 = -2.667\cdot 10^{-4} \;,\quad \tilde{Q}_3 = 3.733\cdot 10^{-4} \;,\quad \tilde{Q}_5 = -1.180\cdot 10^{-3}\;,\quad \tilde{Q}_7 = 6.771\cdot 10^{-3}\;, \notag \\
    & \tilde{Q}_9 = -6.191\cdot 10^{-2}\;, \quad \tilde{Q}_{11} = 8.285\cdot 10^{-1} \;,\quad \tilde{Q}_{13}= -1.528\cdot 10^1 \;, \quad \tilde{Q}_{15} = 3.715 \cdot 10^2 \;, \notag\\
    & \tilde{Q}_{17} = -1.152 \cdot 10^4\;,\quad \tilde{Q}_{19} = 4.434 \cdot 10^5 \;,\quad \tilde{Q}_{21} = -2.075 \cdot 10^7\;,\quad \ldots \;.
\end{align}
Just as for the KdV equation, we see that a small deformation of the initial conditions produces a seemingly arbitrarily large change in the conserved charges. Again, we argue that this behavior is a consequence of the instability of the inverse scattering method. A related way to see this is to notice that the conserved charges can be obtained from the expansion of an analytic function -- the \emph{transfer matrix} -- about one of its singularities\cite{babelon2003introduction, candu2013introduction}, see Appendix \ref{app:mon_mat} for a quick review. The rigidity of analytic functions makes such a procedure very sensitive to small deformations.

\begin{figure}
    \centering
    \includegraphics[scale=0.6]{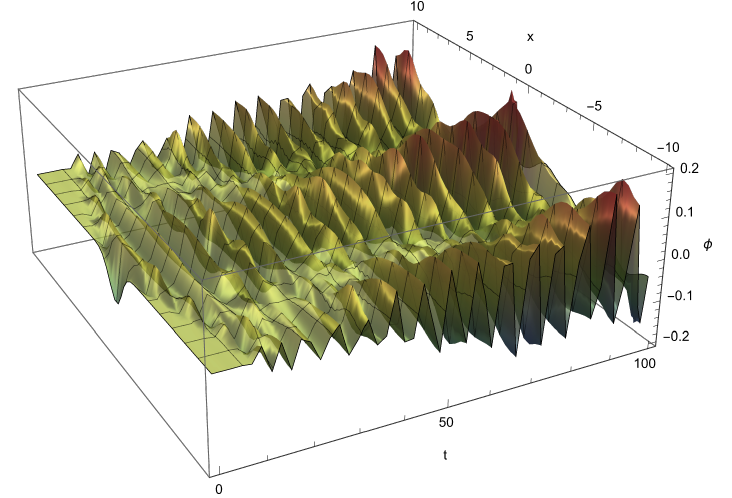}\quad
   \includegraphics[scale=0.8]{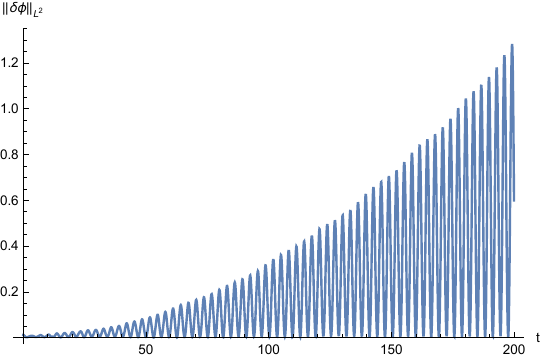}
    \caption{Left: A plot of the difference between the solutions $\phi_1$ and $\phi_2$ to the Cauchy problems (\ref{eq:numdev1}, \ref{eq:numdev2}). Right: the $L_2$-norm of the difference between two solutions as a function of time. The difference grows polynomially with time suggesting that small perturbations can exhibit the {\it weak butterfly effect}.}
    \label{fig:dif}
\end{figure}

As a further evidence suggesting the presence of deterministic chaos in integrable systems, we conduct the following numerical experiment. We consider the sine-Gordon equation, this time on a cylinder of circumference $L$, and compare the evolution of the following two Cauchy problems
\begin{align}
    &\left\{
    \begin{array}{l r}
        \left(\partial_t^2 - \partial_x^2\right) \phi_1(t,x) = \sin \phi_1(t,x)&  \quad x+L \sim x,\\
        \partial_t \phi_1 (0,x) = 0 &  \\ \phi_1(0,x) = \cos \frac{2\pi x}{L} & 
    \end{array}\right.\;, \label{eq:numdev1} \\
    &\left\{
    \begin{array}{l r}
        \left(\partial_t^2 - \partial_x^2\right) \phi_2(t,x) = \sin \phi_2(t,x)&  \quad x+L \sim x, \\
        \partial_t \phi_2 (0,x) = 0 &  \\ \phi_2(0,x) = \cos \frac{2\pi x}{L} + \epsilon \vartheta\left(\frac{x}{L}\right) &
    \end{array}\right.\;,
\label{eq:numdev2}
\end{align}
where we introduced $\vartheta(z) = \sum\limits^\infty_{n=-\infty} \exp\left(-(x-n)^2\right)$, and an arbitrary small parameter $\epsilon\ll1$ . In Figure \ref{fig:dif} we plotted the difference $\phi_1 - \phi_2$. We can see that this quantity grows larger -- in absolute value -- with time and the solutions become increasingly different.

\section{Conclusions and Outlook}
\label{sec:conc_out}
In this note we have presented evidence of an apparent  contradiction to the common lore that integrable systems are incompatible with chaos. In particular we have shown that the map from the initial conditions to the set of conserved charges is unstable, in the sense that can map small differences to arbitrarily large ones. This happens only for the case where the conserved charges are infinite in number, corresponding to systems with an infinite-dimensional phase space. 

There are several open questions that deserve further investigation. Amongst these we highlight the following:
\begin{itemize}
    \item[--] It would be interesting to explore and classify the types of small deformations of integrable systems of the kind explored in this note. In particular identify which of these do not yield an unstable map between the initial conditions and the conserved charges.
    \item[--] In this note we have considered the interplay between chaotic behavior and integrability in the context of classical systems. Naturally one would like to explore this for quantum systems as well.
    \item[--] Recently, an integrable $N$-body system called the \emph{zigzag model} was proposed \cite{Donahue:2019adv,Donahue:2019fgn,Donahue:2022jxu} as a high-energy description of long confining strings in massive adjoint 2D QCD. This model possesses a non-differentiable phase space containing boundaries and distinct topological sectors. In light of the discussion in \S \ref{subsec:FDIS} a natural question concerning the presence of chaos in the zigzag model arises.
    \item[--] Other classical models that have a non-differentiable phase spaces and in which it will be interesting to probe for the existence of chaotic features are the $\mathrm{T}\overline{\mathrm{T}}$-deformed theories \cite{Smirnov:2016lqw,Cavaglia:2016oda} and their generalizations to ``higher $\mathrm{T}\overline{\mathrm{T}}$'' deformations \cite{Conti:2019dxg,Hernandez-Chifflet:2019sua,Camilo:2021gro,Cordova:2021fnr} and to quantum-mechanical $\mathrm{T}\overline{\mathrm{T}}$-like deformations \cite{Gross:2019ach}.
    \item[--] Free theories in any space-time  dimension also admit an infinite number of conserved charges. An interesting question is whether the map between these and the initial conditions exhibit some chaotic feature also for these systems.
    \item[--] It is well known that ``ordinary chaotic behavior" characterizes physical systems that are described by integrable theories perturbed by integrability breaking terms. Examples of these type of theories are the massless Schwinger model perturbed by a mass term and the sine-Gordon model with an additional $\phi^2$ term. The relation between this type of chaos and the one discussed in this work may shed new insight on the concept of chaos in general.
\end{itemize}
\section{Acknowledgement}
The work of S.N. is partially supported by the NSF grant PHY-2210349 and by the Simons Collaboration on Confinement and QCD Strings.  S.N. wishes to thank P. Dorey, R. Tateo and A. Zamolodchikov for precious discussions and suggestions, and the department of physics of the Universit\`{a} degli Studi di Torino for kind hospitality.

F.K.P. is currently a Simons Junior Fellow at NYU and supported by a grant 855325FP from the Simons Foundation.

The work of J.S. was supported in part by the by a center of excellence of the Israel Science Foundation (grant number 2289/18). J.S.  would like to thank M. Bianchi, M. Firrota and D. Weissman for useful  discussions. This work was carried out while J.S. stayed  in NYU and the Simon's Center. He would like to thank both institutes for a warm hospitality.

We are grateful as well to A. Dymarsky, V. Rosenhaus, A. Gorsky, B. Harrop-Griffiths for insightful discussions and suggestions throughout the project.
\appendix
\section{Perturbation Theory for Scattering Data}
\label{app:pert_scatt}
Here we wish to determine how the scattering data varies for small deformations of the potential. Let us then consider the following $1$-dimensional scattering problem
\begin{equation}
    \Big(-\partial_x^2 + u(x)\Big) \Phi(x) = k^2 \Phi(x)\;,
\end{equation}
and two solutions $\phi_k$ and $\psi_k$, satisfying the following boundary conditions
\begin{align}
    &\phi_k(x) \sim \left\{
    \begin{array}{l l}
    e^{-i k x} & x \to -\infty, \\
    \frac{1}{t_k} e^{- i k x} + \frac{r_k}{t_k} e^{i k x} & x \to +\infty
    \end{array}
    \right. \;,\\ &\psi_k(x) \sim 
    e^{-i k x}\;,\quad x \to +\infty \;. \notag
\label{eq:KdVscatteringboundapp}
\end{align}
We wish to study how the transmission and reflection coefficients $t_k$ and $r_k$ vary when we perturb the potential $u(x)$ by a small term $\delta u(x)$. The solution will vary as $\Phi(x)\,\to\,\Phi(x) + \delta\Phi(x)$, where
\begin{equation}
    \delta\Phi(x) = - \intop dy\,G(x,y)\delta u(y)\Phi(y)\;.
\end{equation}
Here $G(x,y)$ is the Green's function for the operator $-\partial_x^2 - k^2 + u(x)$. There exist a unique such function decaying exponentially at large $x$ and $y$ for $\mathrm{Im}k>0$ and it is given by the following expression \cite{feinberg2004response}:
\begin{equation}
    G(x,y) = \frac{i t_k}{2k}\left[\theta(x-y)\psi^\ast_k(x)\phi_k(y) + \theta(y-x)\phi_k(x)\psi^\ast_k(y)\right]\;.
\end{equation}
Hence the variation $\delta\Phi(x)$ will be given by
\begin{equation}
    \delta\Phi(x) = -\frac{i t_k}{2k}\left[\psi_k^\ast(x)\intop_{-\infty}^x dy\,\phi_k(y)\delta u(y)\Phi(y) + \phi_k(x)\intop_x^{\infty} dy\, \psi_k^\ast(y)\delta u(y)\Phi(y)\right]\;.
\end{equation}
We then see that the following function
\begin{equation}
    \tilde{\phi}_k(x) = \frac{\phi_k(x) + \delta\Phi(x)}{1 - \frac{i t_k}{2 k} \intop_{-\infty}^{\infty} dy\,\psi_k^\ast(y)\delta u(y)\Phi_k(y)}\;,
\end{equation}
tends to $e^{-i k x}$ as $x\to-\infty$ and hence constitutes the ``$\phi_k$ solution'' corresponding to the potential $u(x)+\delta u(x)$. We then study its $x\to\infty$ behavior which has the expected form
\begin{equation}
    \tilde{\phi}_k(x) \underset{x\to\infty}{\sim} \frac{1}{\tilde{t}_k} e^{-i k x} + \frac{\tilde{r}_k}{\tilde{t}_k} e^{i k x}\;.
\end{equation}
Using the limiting expressions of $\phi_k(x)$ and $\delta\Phi(x)$, we arrive at the following first order expressions
\begin{align}
    &\delta t_k = \tilde{t}_k - t_k = \frac{ t_k^2}{2ik} \intop_{-\infty}^{\infty} dy\,\psi_k^\ast(y)\delta u(y)\Phi(y)\;, \\
    &\delta r_k = \tilde{r}_k - r_k = \frac{t_k^2}{2ik} \intop_{-\infty}^{\infty} dy\,\phi_k(y)\delta u(y)\Phi(y)\;.
\end{align}
We see that the correction to the scattering data caused by a small perturbation $\delta u(x)$ of the potential receives arbitrarily large contributions in the IR region $k\sim 0$ of the spectrum.

\section{The Transfer Matrix and the Conserved Charges}
\label{app:mon_mat}
Let us consider $2$-dimensional integrable field theories that possess a Lax representation. What this means is that it is possible to represent the equations of motion $\text{EoM}(\phi)$ of the integrable system as a zero curvature condition for a pair of matrices $U$ and $V$. These matrices depend on the space-time variables through the fields $\phi$ and their derivatives. Additionally they depend \emph{analytically} on a spectral parameter $\lambda$, having isolated singularities at fixed positions $\lambda_k$. The zero curvature condition reads
\begin{gather}
    \partial_t U(\phi;\lambda) - \partial_x V(\phi;\lambda) + \big[U(\phi;\lambda),V(\phi;\lambda)\big] = 0\quad \Longleftrightarrow \quad \text{EoM}(\phi)\;,
\label{eq:ZCC}
\end{gather}
and can be interpreted as the consistency condition for the following linear differential system
\begin{gather}
    \partial_x\Psi(x,t;\lambda) = U(\phi;\lambda) \Psi(x,t;\lambda)  \;,\quad \partial_t\Psi(x,t;\lambda) = V(\phi;\lambda) \Psi(x,t;\lambda) \;.
\label{eq:aux_system}
\end{gather}
Fixing an initial condition which is independent on $\lambda$, e.g. $\Psi(0,0;\lambda) = \mathbbm{1}$, we can state that the solution $\Psi(x,t;\lambda)$ is an analytic function of the spectral parameter, with essential singularities at the positions $\lambda = \lambda_k$. The solution to this linear system can be expressed as a path-ordered exponential
\begin{gather}
    \Psi\left(x, t\right) = \operatorname{Pexp}\left( \int_\gamma (U dx + V dt) \right)\Psi_0\;,
\end{gather}
where $\gamma$ is a curve from the origin to the point $(x,t)$. Since the zero curvature condition \eqref{eq:ZCC} is satisfied, the solution $\Psi$ only depends on the endpoints of the path $\gamma$. Choosing $\gamma$ to be a spatial cycle\footnote{For a theory on a cylinder $x\sim x+2\pi$ this will be the path $x\in[0,2\pi)$, while for a theory on the real line it will be the real axis $x\in\mathbb{R}$. Of course, in the latter case, all the functions involved must obey appropriate asymptotic behaviors at $x\rightarrow\pm\infty$ \cite{faddeev1987hamiltonian}.} $\mathscr{C}$ and taking a trace, we arrive at what is known as the \emph{transfer matrix}
\begin{gather}
    \mathcal{T}(\lambda) = \operatorname{Tr} \operatorname{Pexp}\left( \int_{\mathscr{C}} U(\phi;\lambda) dx \right)\;.
\end{gather}
This quantity serves as a generating function for the conserved charges of the theory which can be obtained by expansion about any point. Usually an expansion around a regular point yields the so-called \emph{non-local integrals of motion}. On the other hand, an expansion about one of the singularities $\lambda = \lambda_k$ generates the \emph{local conserved charges}. Let us quickly review the procedure to find the latter for the case of $U$ being a $2\times2$ trace-less matrix having a simple pole\footnote{This is the case for e.g. KdV and sine-Gordon equations. In these cases the pole is located at $\lambda = \infty$.} at $\lambda = \lambda_0$. We further suppose that the diagonal components $U_{11}$ and $-U_{11}$ are regular at $\lambda = \lambda_0$. First it is convenient to perform a \emph{gauge transformation}
\begin{equation}
    U(\phi;\lambda)\;\longrightarrow U^g(\phi;\lambda) = \partial_x g(\phi;\lambda)\,g(\phi;\lambda)^{-1} + g(\phi;\lambda) U(\phi;\lambda) g(\phi;\lambda)^{-1}\;.
\end{equation}
It can be checked that such a transformation leaves the transfer matrix $\mathcal{T}(\lambda)$ invariant. Choosing
\begin{equation}
    g(\phi;\lambda) = \left(\begin{array}{c c}
        e^{-\intop^x U_{11}(\phi;\lambda)} & 0 \\
        0 & e^{\intop^x U_{11}(\phi;\lambda)}
    \end{array}\right)\;,
\end{equation}
produces
\begin{equation}
    U^g(\phi;\lambda) = \left(\begin{array}{c c}
        0 & u(\phi;\lambda) \\
        \tilde u(\phi;\lambda) & 0
    \end{array}\right)\;,\qquad \left\lbrace\begin{array}{l}
        u(\phi;\lambda) = e^{-2\intop^x U_{11}(\phi;\lambda)}U_{12}(\phi;\lambda) \\
        \tilde u(\phi;\lambda) = e^{2\intop^x U_{11}(\phi;\lambda)}U_{21}(\phi;\lambda)
    \end{array}\right.
\end{equation}
Now the $x$ equation of \eqref{eq:aux_system} can be easily expanded in components $\Psi = \left(\psi\;\tilde{\psi}\right)$ as
\begin{equation}
    \partial_x^2\psi - \partial_x\log u\,\partial_x\psi - u\tilde u\psi = 0\;,\qquad \partial_x^2\tilde\psi - \partial_x\log \tilde u\,\partial_x\tilde \psi - u\tilde u\tilde\psi = 0\;.
\end{equation}
One more substitution $\psi = \partial_x\log p$ and $\tilde\psi = \partial_x\log\tilde p$ brings us to the pair of Riccati equations
\begin{equation}
    \partial_x p + p^2 - \partial\log u\,p - u \tilde u = 0\;,\qquad \partial_x \tilde p + \tilde p^2 - \partial\log \tilde u\,\tilde p - u \tilde u = 0\;.
\label{eq:WKB}
\end{equation}
Expanding around $\lambda = \lambda_0$ we find that
\begin{equation}
    p = \sum_{k=-1}^{\infty}p_k\left(\lambda - \lambda_0\right)^k\;,\qquad \left\lbrace \begin{array}{l}
        u = \sum_{k=-1}^{\infty}u_k\left(\lambda - \lambda_0\right)^k \\
        \tilde u = \sum_{k=-1}^{\infty}\tilde u_k\left(\lambda - \lambda_0\right)^k
    \end{array}\right. \;,
\end{equation}
where the coefficients $p_k$ can be found by matching the powers of $\lambda-\lambda_0$ in the equation \eqref{eq:WKB}. For example
\begin{equation}
    p_{-1}^2 = u_{-1}\tilde{u}_{-1}\;,\qquad p_0 = \frac{u_0\tilde{u}_{-1} + u_{-1}\tilde{u}_0}{2p_{-1}} + \frac{1}{2}\partial_x\log\frac{u_{-1}}{p_{-1}}\;.
\end{equation}
Similar equations hold for $\tilde{p}_k$. From the definition of $p$ and $\tilde{p}$, we find that
\begin{equation}
    \psi(x;\lambda) = e^{\intop_{x_0}^x\,p(\xi;\lambda)d\xi}\psi(x_0)\;,\qquad \tilde\psi(x;\lambda) = e^{\intop_{x_0}^x\,\tilde p(\xi;\lambda)d\xi}\tilde\psi(x_0)\;.
\end{equation}
We then conclude that the transfer matrix has the following form
\begin{equation}
    \mathcal{T}(\lambda) = e^{\intop_{\mathscr{C}} p(x;\lambda)dx} + e^{\intop_{\mathscr{C}} \tilde{p}(x;\lambda)dx}\;.
\label{eq:tran_mat_sum}
\end{equation}
Since $\tilde{p}_{-1} = p_{-1}$, the logarithm of the above quantity can be expanded in Laurent series about $\lambda = \lambda_0$
\begin{equation}
    \log\mathcal{T}(\lambda) = \sum_{k=-1}^{\infty} Q_k\left(\lambda - \lambda_0\right)^k\;.
\end{equation}
The quantities $Q_k$ are integrals of motion, e.g.
\begin{equation}
    Q_{-1} = \intop_{\mathscr{C}} p_{-1}(x) dx\;,\qquad Q_{0} = \log\left(e^{\intop_{\mathscr{C}} p_0(x) dx} + e^{\intop_{\mathscr{C}} \tilde{p}_0(x) dx}\right)\;.
\end{equation}
An alternative but equivalent set of integrals of motion $I_k$ can be obtained by expanding directly one of the exponents in \eqref{eq:tran_mat_sum}
\begin{equation}
    \intop_{\mathscr{C}} p(x;\lambda)dx = \sum_{k=-1}^{\infty} I_k \left(\lambda - \lambda_0\right)^k\;.
\end{equation}
The integrals of motion generated by $\tilde{p}$ will not be algebraically independent from the above ones, given the relation $p\tilde{p} = u\tilde{u}$.

\bibliographystyle{ssg}
\bibliography{Chaos}

\begingroup\raggedright\begin{thebibliography}{10}

\bibitem{babelon2003introduction}
O.~Babelon, D.~Bernard, and M.~Talon, {\em {Introduction to Classical
  Integrable Systems}}.
\newblock Cambridge Monographs on Mathematical Physics. Cambridge University
  Press, 2003.

\bibitem{baxter2016exactly}
R.~J. Baxter, {\em {Exactly solved models in statistical mechanics}}.
\newblock 1982.

\bibitem{Dorey:1996gd}
P.~Dorey, ``{Exact S matrices},'' in {\em {Conformal field theories and
  integrable models. Proceedings, Eotvos Graduate Course, Budapest, Hungary,
  August 13-18, 1996}}, pp.~85--125, 1996.
\newblock \href{https://arxiv.org/abs/hep-th/9810026}{{\tt hep-th/9810026}}.

\bibitem{Bombardelli:2016rwb}
D.~Bombardelli, A.~Cagnazzo, R.~Frassek, F.~Levkovich-Maslyuk, F.~Loebbert,
  S.~Negro, I.~M. Szécsényi, A.~Sfondrini, S.~J. van Tongeren, and
  A.~Torrielli, ``{An integrability primer for the gauge-gravity
  correspondence: An introduction},'' {\em J. Phys.} {\bf A49} (2016), no.~32
  320301, \href{https://arxiv.org/abs/1606.02945}{{\tt 1606.02945}}.

\bibitem{Bazhanov:1994ft}
V.~V. Bazhanov, S.~L. Lukyanov, and A.~B. Zamolodchikov, ``Integrable structure
  of conformal field theory, quantum {K}d{V} theory and thermodynamic {B}ethe
  ansatz,'' {\em Commun. Math. Phys.} {\bf 177} (1996) 381--398,
  \href{https://arxiv.org/abs/hep-th/9412229}{{\tt hep-th/9412229}}.

\bibitem{Bazhanov:1996aq}
V.~V. Bazhanov, S.~L. Lukyanov, and A.~B. Zamolodchikov, ``Integrable quantum
  field theories in finite volume: Excited state energies,'' {\em Nucl. Phys.}
  {\bf B489} (1997) 487--531, \href{https://arxiv.org/abs/hep-th/9607099}{{\tt
  hep-th/9607099}}.

\bibitem{Bazhanov:1996dr}
V.~V. Bazhanov, S.~L. Lukyanov, and A.~B. Zamolodchikov, ``Integrable structure
  of conformal field theory. 2. {Q} operator and {DDV} equation,'' {\em Commun.
  Math. Phys.} {\bf 190} (1997) 247--278,
  \href{https://arxiv.org/abs/hep-th/9604044}{{\tt hep-th/9604044}}.

\bibitem{Negro:2016yuu}
S.~Negro, ``{Integrable structures in quantum field theory},'' {\em J. Phys. A}
  {\bf 49} (2016), no.~32 323006, \href{https://arxiv.org/abs/1606.02952}{{\tt
  1606.02952}}.

\bibitem{maldacena2016remarks}
J.~Maldacena and D.~Stanford, ``{Remarks on the Sachdev-Ye-Kitaev model},''
  {\em Phys. Rev. D} {\bf 94} (2016), no.~10 106002,
  \href{https://arxiv.org/abs/1604.07818}{{\tt 1604.07818}}.

\bibitem{maldacena2016bound}
J.~Maldacena, S.~H. Shenker, and D.~Stanford, ``{A bound on chaos},'' {\em
  JHEP} {\bf 08} (2016) 106, \href{https://arxiv.org/abs/1503.01409}{{\tt
  1503.01409}}.

\bibitem{gross2021chaotic}
D.~J. Gross and V.~Rosenhaus, ``{Chaotic scattering of highly excited
  strings},'' {\em JHEP} {\bf 05} (2021) 048,
  \href{https://arxiv.org/abs/2103.15301}{{\tt 2103.15301}}.

\bibitem{deutsch1991quantum}
J.~M. Deutsch, ``Quantum statistical mechanics in a closed system,'' {\em
  Physical review a} {\bf 43} (1991), no.~4 2046.

\bibitem{Motamarri:2021zwf}
V.~Motamarri, A.~S. Gorsky, and I.~M. Khaymovich, ``{Localization and
  fractality in disordered Russian Doll model},''
  \href{https://arxiv.org/abs/2112.05066}{{\tt 2112.05066}}.

\bibitem{guckenheimer2013nonlinear}
J.~Guckenheimer and P.~Holmes, {\em Nonlinear oscillations, dynamical systems,
  and bifurcations of vector fields}, vol.~42.
\newblock Springer Science \& Business Media, 2013.

\bibitem{PlatoStanford}
``Stanford encyclopedia of philosophy, chaos.''
  \url{https://plato.stanford.edu/entries/chaos/}.
\newblock Accessed: 2010-09-30.

\bibitem{zakharov1991integrability}
V.~E. Zakharov {\em et.~al.}, {\em What is integrability?}, vol.~129.
\newblock Springer, 1991.

\bibitem{hitchin2013integrable}
N.~J. Hitchin, G.~B. Segal, and R.~S. Ward, {\em Integrable systems: Twistors,
  loop groups, and Riemann surfaces}, vol.~4.
\newblock OUP Oxford, 2013.

\bibitem{Torrielli:2016ufi}
A.~Torrielli, ``{Lectures on Classical Integrability},'' {\em J. Phys. A} {\bf
  49} (2016), no.~32 323001, \href{https://arxiv.org/abs/1606.02946}{{\tt
  1606.02946}}.

\bibitem{zheng2003observer}
Z.~Zheng, B.~Misra, and H.~Atmanspacher, ``Observer-dependence of chaos under
  Lorentz and Rindler transformations,'' {\em International Journal of
  Theoretical Physics} {\bf 42} (2003), no.~4 869--879.

\bibitem{zakharov1980inverse}
V.~Zakharov, ``The inverse scattering method,'' in {\em Solitons},
  pp.~243--285.
\newblock Springer, 1980.

\bibitem{faddeev1987hamiltonian}
L.~D. Faddeev and L.~A. Takhtajan, {\em Hamiltonian methods in the theory of
  solitons}, vol.~23.
\newblock Springer, 1987.

\bibitem{smilansky1991classical}
U.~Smilansky, ``The classical and quantum theory of chaotic scattering,'' {\em
  Chaos et Physique Quantique, Les Houches LII, M.-J. Giannoni, A. Voros and J.
  Zinn-Justin eds., Elsevier} (1991) 371--441.

\bibitem{eckhardt1987fractal}
B.~Eckhardt, ``Fractal properties of scattering singularities,'' {\em Journal
  of Physics A: Mathematical and General} {\bf 20} (1987), no.~17 5971.

\bibitem{gel1951determination}
I.~M. Gel'fand and B.~M. Levitan, ``On the determination of a differential
  equation from its spectral function,'' {\em Izvestiya Rossiiskoi Akademii
  Nauk. Seriya Matematicheskaya} {\bf 15} (1951), no.~4 309--360.

\bibitem{marchenko2011sturm}
V.~A. Marchenko, {\em Sturm-Liouville operators and applications}, vol.~373.
\newblock American Mathematical Soc., 2011.

\bibitem{dorren1994stability}
H.~Dorren, E.~Muyzert, and R.~Snieder, ``The stability of one-dimensional
  inverse scattering,'' {\em Inverse Problems} {\bf 10} (1994), no.~4 865.

\bibitem{carrion1986stability}
P.~Carrion, ``On stability of 1D exact inverse methods,'' {\em Inverse
  Problems} {\bf 2} (1986), no.~1.

\bibitem{feinberg2004response}
J.~Feinberg, ``The response to a perturbation in the reflection amplitude,''
  {\em Journal of Physics A: Mathematical and General} {\bf 37} (2004), no.~43
  10261.

\bibitem{berryman1980discrete}
J.~G. Berryman and R.~R. Greene, ``Discrete inverse methods for elastic waves
  in layered media,'' {\em Geophysics} {\bf 45} (1980), no.~2 213--233.

\bibitem{landau2013quantum}
L.~D. Landau and E.~M. Lifshitz, {\em Quantum mechanics: non-relativistic
  theory}, vol.~3.
\newblock Elsevier, 2013.

\bibitem{candu2013introduction}
C.~Candu and M.~de~Leeuw, ``Introduction to integrability,'' {\em lectures
  delivered at ETH} (2013).

\bibitem{Donahue:2019adv}
J.~C. Donahue and S.~Dubovsky, ``{Confining Strings, Infinite Statistics and
  Integrability},'' {\em Phys. Rev. D} {\bf 101} (2020), no.~8 081901,
  \href{https://arxiv.org/abs/1907.07799}{{\tt 1907.07799}}.

\bibitem{Donahue:2019fgn}
J.~C. Donahue and S.~Dubovsky, ``{Classical Integrability of the Zigzag
  Model},'' {\em Phys. Rev. D} {\bf 102} (2020), no.~2 026005,
  \href{https://arxiv.org/abs/1912.08885}{{\tt 1912.08885}}.

\bibitem{Donahue:2022jxu}
J.~C. Donahue and S.~Dubovsky, ``{Quantization of the zigzag model},'' {\em
  JHEP} {\bf 08} (2022) 047, \href{https://arxiv.org/abs/2202.11746}{{\tt
  2202.11746}}.

\bibitem{Smirnov:2016lqw}
F.~A. Smirnov and A.~B. Zamolodchikov, ``{On space of integrable quantum field
  theories},'' {\em Nucl. Phys. B} {\bf 915} (2017) 363--383,
  \href{https://arxiv.org/abs/1608.05499}{{\tt 1608.05499}}.

\bibitem{Cavaglia:2016oda}
A.~Cavagli\`a, S.~Negro, I.~M. Sz\'ecs\'enyi, and R.~Tateo, ``{$T
  \bar{T}$-deformed 2D Quantum Field Theories},'' {\em JHEP} {\bf 10} (2016)
  112, \href{https://arxiv.org/abs/1608.05534}{{\tt 1608.05534}}.

\bibitem{Conti:2019dxg}
R.~Conti, S.~Negro, and R.~Tateo, ``{Conserved currents and
  $\text{T}\bar{\text{T}}_s$ irrelevant deformations of 2D integrable field
  theories},'' {\em JHEP} {\bf 11} (2019) 120,
  \href{https://arxiv.org/abs/1904.09141}{{\tt 1904.09141}}.

\bibitem{Hernandez-Chifflet:2019sua}
G.~Hern\'andez-Chifflet, S.~Negro, and A.~Sfondrini, ``{Flow Equations for
  Generalized $T\overline{T}$ Deformations},'' {\em Phys. Rev. Lett.} {\bf 124}
  (2020), no.~20 200601, \href{https://arxiv.org/abs/1911.12233}{{\tt
  1911.12233}}.

\bibitem{Camilo:2021gro}
G.~Camilo, T.~Fleury, M.~Lencs\'es, S.~Negro, and A.~Zamolodchikov, ``{On
  factorizable S-matrices, generalized TTbar, and the Hagedorn transition},''
  {\em JHEP} {\bf 10} (2021) 062, \href{https://arxiv.org/abs/2106.11999}{{\tt
  2106.11999}}.

\bibitem{Cordova:2021fnr}
L.~C\'ordova, S.~Negro, and F.~I. Schaposnik~Massolo, ``{Thermodynamic Bethe
  Ansatz past turning points: the (elliptic) sinh-Gordon model},'' {\em JHEP}
  {\bf 01} (2022) 035, \href{https://arxiv.org/abs/2110.14666}{{\tt
  2110.14666}}.

\bibitem{Gross:2019ach}
D.~J. Gross, J.~Kruthoff, A.~Rolph, and E.~Shaghoulian, ``{$T\overline{T}$ in
  AdS$_2$ and Quantum Mechanics},'' {\em Phys. Rev. D} {\bf 101} (2020), no.~2
  026011, \href{https://arxiv.org/abs/1907.04873}{{\tt 1907.04873}}.

\end{thebibliography}\endgroup
    
\end{document}